\documentstyle[11pt,newpasp,twoside]{article}
\index{summary}
\index{instructions}
\index{template}

\newcommand{\beq}{\begin{equation}}
\newcommand{\eeq}{\end{equation}}

\def\edcomment#1{\iffalse\marginpar{\raggedright\sl#1\/}\else\relax\fi}
\reversemarginpar

\begin{document}
\title{Modelling Plerion Spectra and their Evolution}
\author{Yves A. Gallant}
\affil{Service d'Astrophysique, C.E.A. Saclay, 91191 Gif-sur-Yvette,
       France}
\author{Eric van der Swaluw}
\affil{D.I.A.S.,
       5 Merrion Square, Dublin 2, Ireland}
\author{John G. Kirk}
\affil{MPI f\"ur Kernphysik,
       Postfach 10\,39\,80, 69029 Heidelberg, Germany}
\author{Abraham Achterberg}
\affil{Astronomical Institute Utrecht, Postbus 80\,000, 3508\,TA Utrecht,
       Netherlands}

\begin{abstract}
We review recent theoretical developments on pulsar winds, their
nebulae and relativistic shock acceleration, and show how they
illuminate unsolved problems in plerion spectra, in particular
the multiple spectral breaks in the Crab and the low-frequency
breaks of plerions such as G\,21.5--0.9 and 3C\,58.
Recent work on Fermi acceleration theory at relativistic shocks
shows that a particle spectral index of 2.2--2.3, compatible with
the X-ray spectra of plerions, results under a wide variety of
assumptions.  If pulsar winds contain ions as well as electrons
and positrons, the mechanism of Hoshino et al.\ (1992), which
yields harder spectra, would operate at lower energies and may
explain the flat radio spectral indices of plerions.  This
scenario implies wind parameters in the Crab compatible with
the pulsar wind acceleration model of Lyubarsky \& Kirk (2001).
Recent hydrodynamical simulations of plerion evolution inside
SNR blast waves demonstrate that the passage of the reverse
shock rapidly compresses the plerion.  Using a simple isobaric
model, we investigate the influence of the resulting magnetic
field compression and decrease in shock radius on the evolution
of the plerion spectrum.  We suggest that the passage of the
reverse shock may explain the low-frequency breaks in 3C\,58 and
G\,21.5--0.9, as well as the increase in 3C\,58's radio flux.
\end{abstract}

\section{X-ray Spectra of Plerions}

\subsection{Observations and Synchrotron Cooling}

   Plerions are characterized in X-rays by hard, nonthermal
power-law spectra.  In the case of the Crab Nebula, where
statistics are best, the total integrated spectrum has a
best-fit power-law index $\alpha_X = 1.1$ (Toor \& Seward
1974) in energy ($F_\nu \propto \nu^{-\alpha}$, corresponding
to a photon index $\Gamma_X \equiv \alpha_X + 1 = 2.1$).
Within their larger uncertainties, the X-ray power-law
indices of most other plerions appear compatible with this
value.

   The recent availability of spatially resolved spectra of
plerions with {\em Chandra} and {\em XMM} reveals spectral
steepening towards the edges, with the hardest spectrum at
the center having an index around $\Gamma_X \approx 1.6$.
The steepening is indicative of synchrotron cooling of a
centrally injected hard power-law distribution of electrons
and positrons, as is the difference of 0.5 between the
central and spatially integrated spectral indices.  The
corresponding injected particle spectral index is $p = 2.2$,
defined by
\beq
   \dot{N}(\gamma) \,{\rm d} \gamma
   \propto \gamma^{-p} \,{\rm d} \gamma
\eeq
where $\dot{N}$ is the injection rate and $\gamma$ the
particle Lorentz factor.

\subsection{Theory: Fermi Acceleration at Ultra-Relativistic Shocks}

   This non-thermal population of electrons and positrons
is generally assumed to be accelerated at the termination
shock of a highly relativistic pulsar wind.  Recent
investigations of Fermi acceleration at such relativistic
shocks (Bednarz \& Ostrowski 1998; Gallant \& Achterberg
1999; Kirk et al.\ 2000; Achterberg et al.\ 2001) have
shown that the resulting spectra, in the limit of high
Lorentz factors and of a turbulent magnetic field downstream,
have power-law indices $p$ in the range 2.2--2.3 for a variety
of transport assumptions, compatible with the above inferred
value for the injected spectrum.

   While further investigations, in particular using 3-D
plasma simulations, are needed to confirm that the required
levels of magnetic turbulence can be achieved, it seems
reasonable to identify the acceleration mechanism for X-ray
emitting electrons with Fermi acceleration at the pulsar
wind termination shock.  Further observational evidence
for this scenario comes from gamma-ray burst afterglows,
whose spectra can also be explained in terms of Fermi
acceleration at the highly relativistic outer blast wave
(e.g. Gallant et al.\ 2000).

\section{Radio Spectra of Plerions}

\subsection{Observations and Broken Power Laws}

   In the radio domain, plerions are characterized by flat
power-law spectral indices, $\alpha_r \approx 0$; the Crab
Nebula, with $\alpha_r = 0.3$, has a steeper than average
spectrum.  The synchrotron loss time 
of radio-emitting electrons is typically much longer than
the age of the plerion, so that the observed spectrum is the
injection spectrum integrated over the history of the plerion.
Nonetheless, with conventional assumptions about plerion
evolution, spectral models assuming a single power-law
injection spectrum (Pacini \& Salvati 1973; Reynolds \&
Chevalier 1984) are unable to reproduce the observed spectral
index differences $\Delta \alpha > 0.5$ between the radio
and X-ray spectra.

   Evidence in the Crab Nebula suggests that radio-emitting
electron acceleration is still taking place at present, in
the central regions near the pulsar wind termination shock
(Gallant \& Tuffs 2000, 2001; Bietenholz et al.\ 2001b).
Combined with the fact that the nebular spectrum joins
smoothly between the radio and X-ray domains, this suggests
the injection of a single population of accelerated particles
with a broken power-law spectrum.

\subsection{Theory: Resonant Ion Cyclotron Wave Acceleration}

   Hoshino et al.\ (1992) have shown that if pulsar winds
contain some ions as well as electrons and positrons, an
efficient mechanism to accelerate the positrons and electrons
is by resonant absorption of ion cyclotron waves collectively
emitted at the shock front.  Particle-in-cell simulations of
this process yielded a range of spectral indices, but these
are in general hard, with an average close to the value of
$p=1.6$ needed to explain the Crab Nebula radio spectrum.
Moreover, in a quasi-stationary calculation of this process
where emission is balanced by absorption (Hoshino \& Arons 1991),
as might be expected in older systems, the resulting spectral
index is $p=1$, yielding a synchrotron spectral index
$\alpha = 0$, exactly the generic plerion radio value.

   This resonant ion cyclotron wave mechanism was seen in the
simulations of Hoshino et al.\ (1992) to accelerate positrons
and electrons up to a critical energy
\beq
\gamma_{\rm crit} \sim \frac{m_i}{m_e} \gamma_{\rm sh}
\label{gcrit}
\eeq
at which they resonate with the fundamental ion cyclotron
frequency; here $m_i / m_e$ is the mass ratio between ions
and electrons, and $\gamma_{\rm sh}$ is the shock Lorentz
factor, which is approximately the downstream thermal
ion Lorentz factor.  From the condition that the electrons
must have gyro-radii larger than the shock thickness for Fermi
acceleration to operate, it follows that $\gamma_{\rm crit}$
is also the minimum energy for the Fermi mechanism.  The
picture that emerges for the accelerated particle spectrum
is thus of a broken power-law, with a hard spectral index
$p_r = 1$--1.6 up to $\gamma_{\rm crit}$, and the steeper
Fermi acceleration index $p_X = 2.2$ at higher energies.

\section{Cooling Break, Injection Break, and Crab Wind Parameters}

   If the X-ray spectrum is synchrotron-cooled and the radio
one is not, they must be separated by a synchrotron cooling
break, with $\Delta\alpha \approx 0.5$, in addition to the
aforementioned injection break.  The Crab Nebula, whose spectrum
is known over most of the intervening frequency range, does indeed
show two spectral breaks, one in the far infrared (FIR) around
$3 \times 10^{13}$\,Hz, and one in the UV around $10^{16}$\,Hz.
(A third break, around 100\,keV, will not be discussed here.)
Spectral index mapping in the IR, where extinction introduces
less uncertainty than in the optical, supports the identification
of the FIR break as the synchrotron cooling break (Gallant \&
Tuffs 2000, 2001).  Given the age of the Nebula, this yields
a magnetic field $\langle B \rangle \approx 3 \times 10^{-4}$\,G,
in line with other estimates of the nebular magnetic field.

   Identification of the injection break with the UV break then
allows inference of the wind Lorentz factor through (\ref{gcrit}).
This yields for the Crab a value of $\gamma_{\rm sh} \sim 10^3$,
in stark contrast to the oft-repeated figure of $10^6$ from
Kennel \& Coroniti (1984).  It should be emphasized, however,
that their model not only does not account for the radio-emitting
electrons and positrons in the Nebula, but is incompatible with
their originating from the pulsar.  If one assumes that all the
radio-emitting $e^\pm$ were injected in the Nebula by the pulsar
wind, it follows that the time-averaged injection rate of pairs must
be $\dot{N}_\pm \sim 3 \times 10^{40}\, {\rm s}^{-1}$, corresponding
with current Crab parameters to a pair multiplicity $\kappa \sim
10^6$.  A wind carrying this number of pairs with a Lorentz factor
of $10^6$ would exceed the Crab spindown power by several orders
of magnitude.

   There are few theoretical predictions of pulsar wind parameters
at the termination shock.
One is given by the reconnecting striped wind model of Lyubarsky
\& Kirk (2001), who derive an asymptotic solution for the wind
Lorentz factor as a function of radius.  Substituting the Crab
pulsar parameters and the above multiplicity of $10^6$ yields a
wind Lorentz factor at the termination shock $\gamma_{\rm sh}
\approx 2 \times 10^3$, in close agreement with our above
observationally derived value.  It should be noted, however,
that the required high multiplicity is problematic for pair
creation models above pulsar polar caps: the recent calculations
of Hibschman \& Arons (2001) yield $\kappa \sim 10^5$ for the
Crab pulsar.

\section{Evolution of Plerion Spectra}

\subsection{Compression by the Reverse Shock}

   An analytical framework for the spectral evolution of an
expanding plerion was presented by Pacini \& Salvati (1973),
and extended by Reynolds \& Chevalier (1984) to the more
realistic case of a plerion evolving inside a shell supernova
remnant (SNR).  Their analyses remain valid in our picture,
except for the addition of a break in the injected spectrum.
In particular, in the initial phase of supersonic expansion
of the plerion, the particle energies and magnetic field both
decrease with time as the plerion radius increases, yielding
a very steep decline of the radio surface brightness with
time, $S_\nu \propto t^{-2.7}$ for $p=1$ and an approximately
constant pulsar spindown luminosity (Reynolds \& Chevalier
1984).  This helps explain the apparent absence of a radio
plerion around some middle-aged pulsars.

   This initial phase lasts until the reverse shock from the
SNR blast wave reaches the central plerion; at that point,
the plerion can be dramatically compressed, as shown by the
recent hydrodynamical simulations of van der Swaluw et al.\
(2001), and subsequently undergoes a slower and initially
unsteady subsonic expansion.  Compression by the reverse
shock increases the magnetic field and particle energies
while the radius decreases, which can lead to a significant
rebrightening of the plerion (Reynolds \& Chevalier 1984).
The increased magnetic field can also bring down the
frequency of the synchrotron cooling break.

\subsection{Low-Frequency Spectral Breaks}

   A number of plerions such as G\,21.5--0.9 and 3C\,58
have observed or inferred spectral breaks at comparatively
low frequencies ($\sim$100\,GHz).  Woltjer et al.\ (1997)
have argued that these form a separate class of plerions,
and can be explained by a sudden change in the particle
or magnetic energy content of the pulsar wind.  Here we
suggest that the sudden change may occur not in the pulsar
wind parameters, but in the plerion confining pressure,
as occurs with the reverse shock passage.  Assuming that
the compression results in a magnetic field in rough
pressure balance with the interior of a Sedov blast wave,
and that the compression lasts for a time comparable to
the age at which the reverse shock hit, one finds that
the synchrotron cooling break can reach the observed low
frequencies only in a very dense surrounding medium
($n \sim 30\, {\rm cm}^{-3}$).

   Such a scenario might be plausible in the case of
G\,21.5--0.9, where it could explain the small size of
the X-ray ``halo'' around the plerion, interpreted as a
non-thermal shell (Slane et al.\ 2000).  The high density
would lead to efficient radiative cooling and a rapid
shock deceleration, which might explain the lack of
observed thermal X-ray emission from this shell.  The
high extinction to this object ($A_V \sim 10$ from the
measured X-ray $N_H$) would make any optical signature
of the shell unobservable.  As for 3C\,58, its apparent
radio brightening with time (Green 1987) may be direct
observational evidence that it is currently undergoing
compression.  This would be consistent with the lack of
detected radio expansion of this remnant (Bietenholz
et al.\ 2001a), but the absence of any detected shell
emission remains a puzzle.

\section{Conclusions}

   The X-ray spectra of plerions are compatible with
Fermi acceleration at ultra-relativistic shocks, which
yields a power-law distribution of injected particles
with spectral index $p_X = 2.2$--2.3, above a critical
energy $\gamma_{\rm crit}$.  Plerion radio spectra
seem compatible with resonant ion cyclotron wave
acceleration, yielding a harder power-law index,
$p_r = 1$--1.6, and fixing the break energy at
$\gamma_{\rm crit} \sim (m_i/m_e) \gamma_{\rm sh}$.
In the case of the Crab Nebula, identification of the
synchrotron cooling break in the FIR and the injection
break in the UV inplies a wind Lorentz factor of about
$10^3$, in sharp contrast with the model of Kennel \&
Coroniti (1984).  This value of the wind Lorentz
factor is compatible with the striped wind model of
Lyubarsky \& Kirk (2001).  Finally, compression by
the reverse shock might be responsible for the
low-frequency breaks observed in G\,21.5--0.9 and
3C\,58, among others, but this requires a dense
surrounding medium.

   One prediction of our scenario for particle
acceleration is that small-scale features near the
wind termination shock, where synchrotron losses
have not had time to operate, should reflect the
unsteepened injection spectrum, with a single
spectral break.  The Crab Nebula's wisps should
be just such features, and comparison of the
frequency of this injection break in the wisps
and the Nebula as a whole would then allow a
determination of the relative magnetic field
values.

\acknowledgments

Y.A.G. is currently supported by a Marie Curie Fellowship from
the European Union, number MCFI-2000-00855.  This work is a
collaboration within the ``Astro-Plasma-Physics'' Network,
supported by the EU under the TMR program, contract number
FMRX-CT98-0168.


\begin{references}

\reference Achterberg, A., Gallant, Y.A., Kirk, J.G., \& Guthmann, A.W.
           2001, \mnras, 328, 393
\reference Bednarz, J., \& Ostrowski, M. 1998, \prl, 80, 3911
\reference Bietenholz, M.F., Kassim, N.E., \& Weiler, K.W. 2001a, \apj,
           560, 772
\reference Bietenholz, M., Bartel, N., Frail, D., \& Hester, J. 2001b,
           these proceedings
\reference Gallant, Y.A., \& Achterberg, A. 1999, \mnras, 305, L6
\reference Gallant, Y.A., Achterberg, A., Kirk, J.G., \& Guthmann, A.W.
           2000, in AIP Conf.\ Ser.\ Vol.\ 526, Gamma-Ray Bursts: 5th
           Huntsville Symposium, ed.\ R.M. Kippen, R.S. Mallozzi \&
           G.J. Fishman (New York: AIP), 524
\reference Gallant, Y.A., \& Tuffs, R.J. 2000, in ASP Conf.\ Ser.\
           Vol.\ 202, Pulsar Astronomy --- 2000 and Beyond, ed.\
           M. Kramer, N. Wex \& N. Wielebinski (San Francisco: ASP),
           503
\reference Gallant, Y.A., \& Tuffs, R.J. 2001, these proceedings
\reference Green, D.A. 1987, \mnras, 225, 11P
\reference Hibschman, J.A. \& Arons, J. 2001, \apj, 554, 624
\reference Hoshino, M., \& Arons, J. 1991, Phys. Fluids B, 3, 818
\reference Hoshino, M., Arons, J., Gallant, Y.A., \& Langdon, A.B. 1992,
           \apj, 390, 454
\reference Kennel, C. F., \& Coroniti, F. V. 1984, \apj, 283, 710
\reference Kirk, J.G., Guthmann, A.W., Gallant, Y.A., \& Achterberg, A.
           2000, \apj, 542, 235
\reference Lyubarsky, Y. \& Kirk, J.G. 2001, \apj, 547, 437
\reference Pacini, F., \& Salvati, M. 1973, \apj, 186, 249
\reference Reynolds, S.P., \& Chevalier, R.A. 1984, \apj, 278, 630
\reference Slane, P., Chen, Y., Schulz, N.S., Seward, F.D., Hughes,
           J.P., \& Gaensler, B.M. 2000, \apj 533, L29
\reference Toor, A., \& Seward, F.D. 1974, \aj, 79, 995
\reference van der Swaluw, E., Achterberg, A., Gallant, Y.A., \&
           T\"oth, G. 2001, \aap, in press (astro-ph/0012440)
\reference Woltjer, L., Salvati, M., Pacini, F., \& Bandiera, R.
           1997, \aap, 325, 295

\end{references}
\end{document}